\documentclass[aps,pra,showpacs,superscriptaddress,preprint]{revtex4}
\usepackage{amsmath}
\usepackage{epsf}
\usepackage{epsfig}

\begin{document}

\title{Time dependence of joint entropy of oscillating quantum systems}

\author{\"{O}zg\"{u}r \"{O}ZCAN}
\email[E-mail: ]{ozcano@hacettepe.edu.tr}\affiliation{Department
of Physics Education, Hacettepe University, 06800, Ankara,Turkey}

\author{Ethem AKT\"{U}RK}
\email[E-mail: ]{eakturk@hacettepe.edu.tr}\affiliation{Department
of Physics, Hacettepe University, 06800, Ankara,Turkey}
\author{Ramazan SEVER}
\email[E-mail: ]{sever@metu.edu.tr}\affiliation{Department of
Physics, Middle East Technical University, 06531, Ankara,Turkey}
\begin{abstract}
The time dependent entropy (or Leipnik's entropy) of harmonic and
damped harmonic oscillators is extensively investigated by using
time dependent wave function obtained by the Feynman path integral
method. Our results for simple harmonic oscillator are in agrement
with the literature. However, the joint entropy of damped harmonic
oscillator shows remarkable discontinuity with time for certain
values of damping factor. According to the results, the envelop of
the joint entropy curve increases with time monotonically. This
results is the general properties of the envelop  of the joint
entropy curve for quantum systems.

\end{abstract}

\maketitle

\section{Introduction}
The investigation of time dependent entropy of the quantum
mechanical systems attracts much attention in recent years. For both
open and closed quantum systems, the different information-theoretic
entropy measures have been discussed
~\cite{Zurek,Omnes,Anastopoulos}. In contrast, the joint
entropy~\cite{Leipnik,Dodonov} can also be used to measure the loss
of information, related to evolving pure quantum
states~\cite{Trigger}. The joint entropy of the physical systems
which are named MACS (maximal classical states) were conjectured by
Dunkel and Trigger~\cite{Dunkel}. According to Ref.~\cite{Dunkel},
the joint entropy of the quantum mechanical systems increase
monotonically with time but this results are not sufficient for
simple harmonic oscillator~\cite{Garbaczewski}.

The aim of this study is to calculate the complete joint entropy
information analytically for simple harmonic and damped harmonic
oscillator systems.

This paper is organized as follows. In section II, we explain
fundamental definitions  needed for the calculations. In section
III, we deal with calculation and results for harmonic oscillator
systems. Moreover, we obtain the analytical solution of Kernel,
wave function in both coordinate and momentum space and its joint
entropy. We also obtain same quantities for damped harmonic
oscillator case. Finally, we present the conclusion in section IV.

\section{Fundamental Definitions}
We deal with a classical system with $d=sN$ degrees of freedom,
where N is the particle number and s is number of spatial
dimensions~\cite{Dunkel}. We assume that the density function
$g(x,p,t)=g(x_1,...,x_d,p_1,...,p_d,t)$ which is the non-negative
time dependent phase space density function of the system has been
normalized to unity,
\begin{equation}
\int dx dp g(x,p,t)=1.
\end{equation}
The Gibbs-Shannon entropy is described by
\begin{equation}
S(t)=-\frac{1}{N!}\int dx dp g(x,p,t)ln(h^{d} g(x,p,t)),
\end{equation}
where $h=2\pi\hbar$ is the Planck constant. Schr\"{o}dinger wave
equation with the Born interpretation~\cite{Born} is given by
\begin{equation}
i\hbar\frac{\partial\psi}{\partial t}=\hat{H}\psi.
\end{equation}
The quantum probability densities are defined in position and
momentum spaces as $|\psi(x,t)|^2$ and $|\tilde{\psi}(p,t)|^2$,
where $|\tilde{\psi}(p,t)|^2$ is given  as
\begin{equation}
    \tilde{\psi}(p,t)=\int\frac{dx
    e^{-ipx/\hbar}}{(2\pi\hbar)^{d/2}}\psi(x,t).
\end{equation}
Leipnik proposed the product function as~\cite{Dunkel}
\begin{equation}
    g_{j}(x,p,t)=|\psi(x,t)|^2|\tilde{\psi}(p,t)|^2\geq0.
\end{equation}
Substituting Eq. (5) into Eq. (2), we get the joint entropy
$S_{j}(t)$ for the pure state $\psi(x,t)$ or equivalently it can
be written in the following form~\cite{Dunkel}
\begin{eqnarray}
    S_{j}(t)&=&-\int dx |\psi(x,t)|^{2}\ln|\psi(x,t)|^{2}-
    \int dp |\tilde{\psi}(p,t)|^2
    \ln |\tilde{\psi}(p,t)|^2-\nonumber\\&-&\ln h^{d}.
\end{eqnarray}
 We find time dependent wave function by means of the Feynman path
integral which has form~\cite{Feynman}
\begin{eqnarray}
K(x'',t'';x',t')&=&\int^{x''=x(t'')}_{x'=x(t')}Dx(t)e^{\frac{i}{\hbar}S[x(t)]}
\nonumber\\&=&\int^{x''}_{x'}Dx(t)e^{\frac{i}{\hbar}\int_{t'}^{t''}L[x,\dot{x},t]dt}.
\end{eqnarray}
The Feynman kernel can be related to the time dependent
Schr\"{o}dinger's wave function
\begin{eqnarray}
K(x'',t'';x',t')=\sum_{n=0}^{\infty}\psi_{n}^{*}(x',t')\psi_{n}(x'',t'').
\end{eqnarray}
The propagator in semiclassical approximation reads
\begin{eqnarray}
K(x'',t'';x',t')=\Big[\frac{i}{2\pi\hbar}\frac{\partial^2}{\partial
x'\partial
x''}S_{cl}(x'',t'';x',t')\Big]^{1/2}e^{\frac{i}{\hbar}S_{cl}(x'',t'';x',t')}.
\end{eqnarray}
The prefactor is often referred to as the Van Vleck-Pauli-Morette
determinant ~\cite{Khandekar,Kleinert}. The $F(x'',t'';x',t')$ is
given by
\begin{eqnarray}
F(x'',t'';x',t')=\Big[\frac{i}{2\pi\hbar}\frac{\partial^2}{\partial
x'\partial x''}S_{cl}(x'',t'';x',t')\Big]^{1/2}.
\end{eqnarray}

\section{CALCULATION AND RESULTS}

\subsection{Simple Harmonic Oscillator (SHO)}
\begin{figure}[htbp]
\centering \epsfig{file=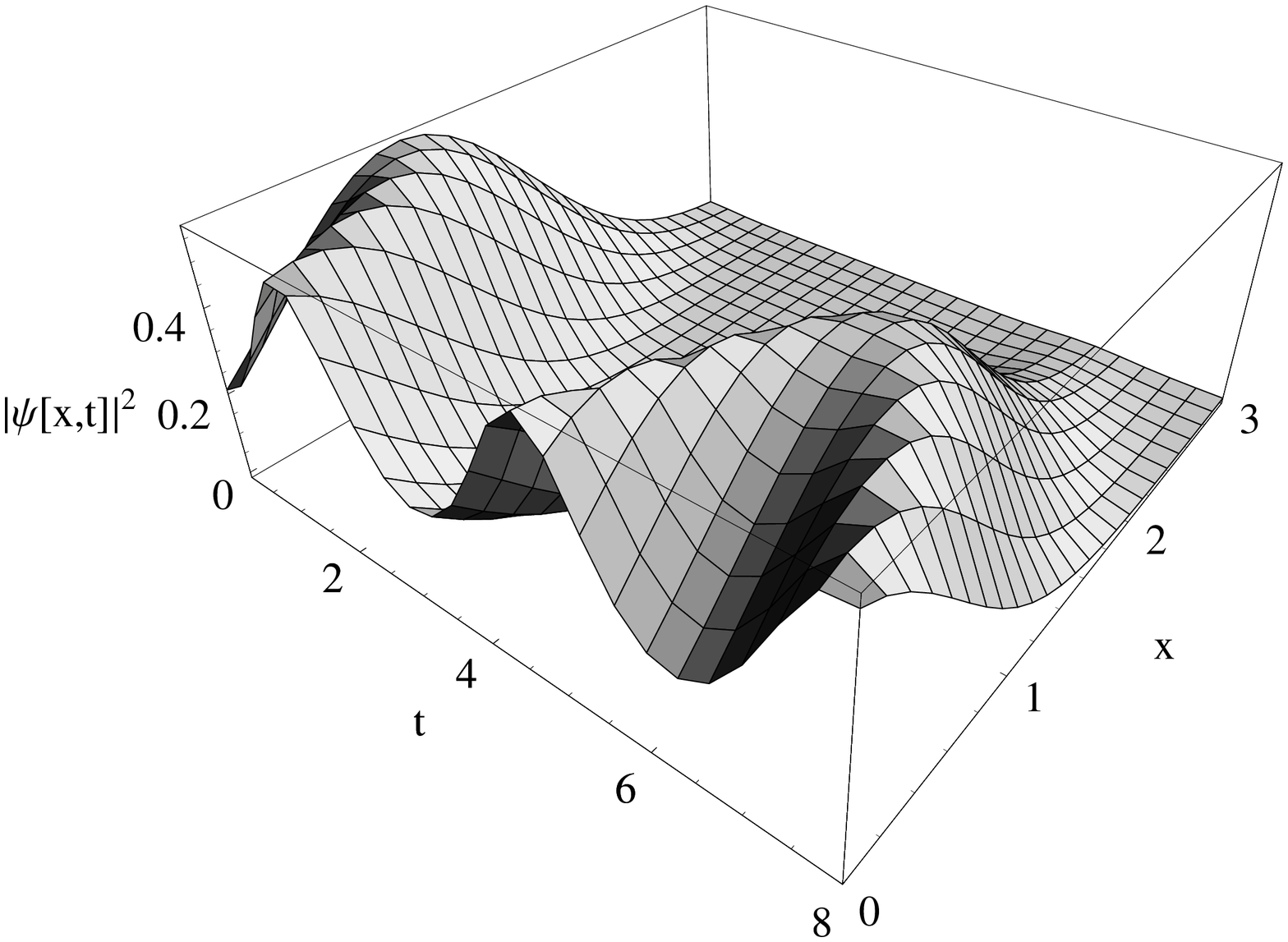, width=12cm,height=12cm}
\caption{$|\Psi(x,t)|^{2}$ versus time and coordinate}\label{eps1}
\end{figure}

\begin{figure}[htbp]
\centering \epsfig{file=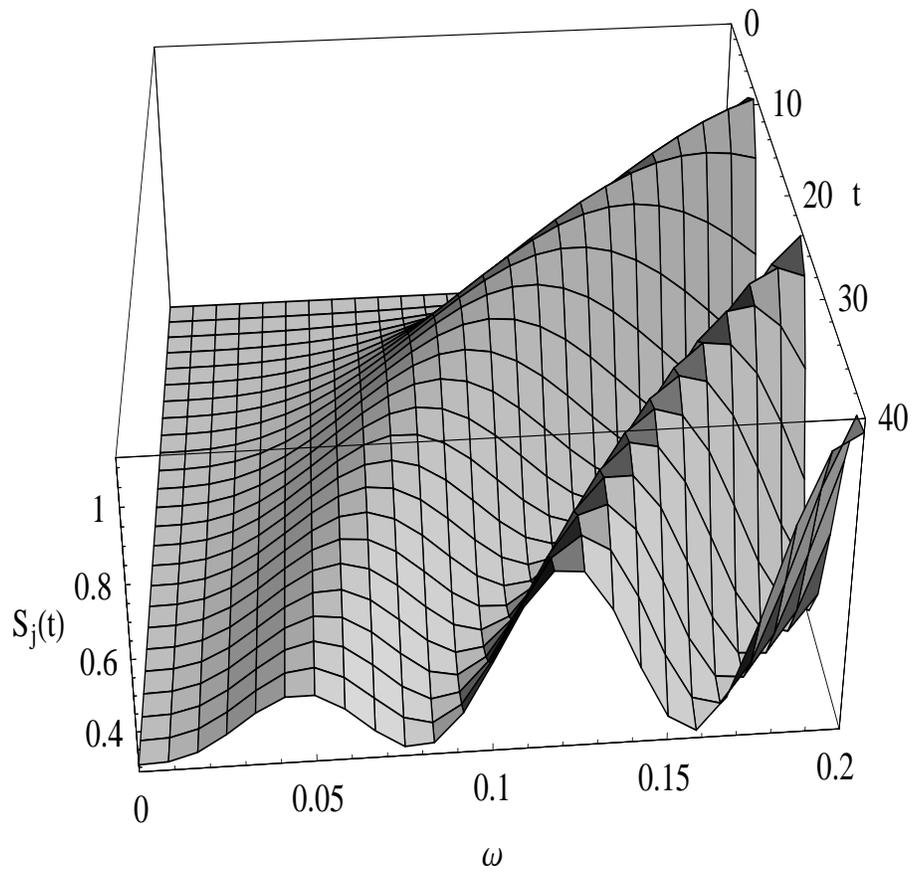, width=12cm,height=12cm}
\caption{The 3D graph of joint entropy of simple harmonic
          oscillator.}\label{eps2}
\end{figure}

\begin{figure}[htbp]
\centering \epsfig{file=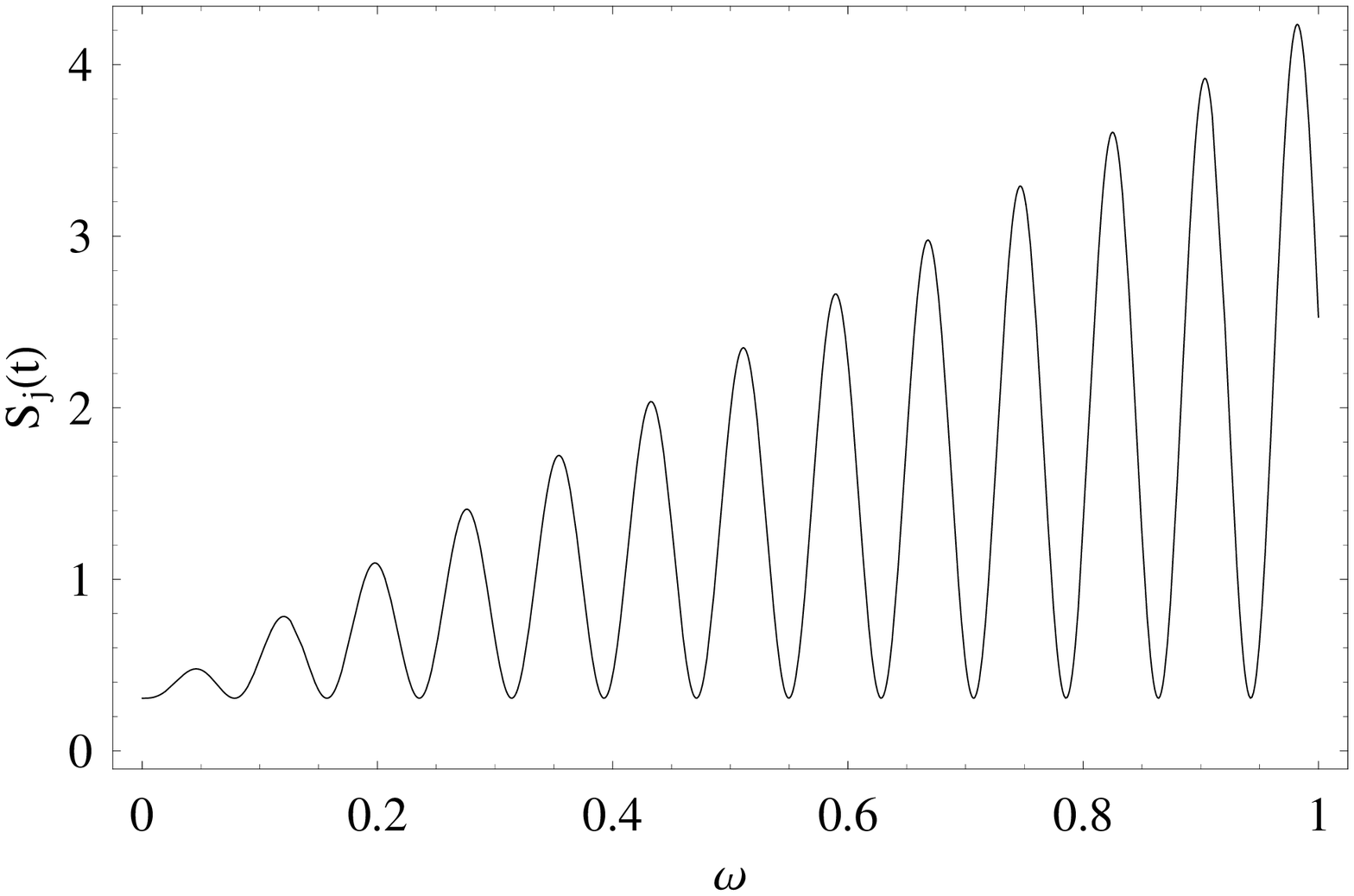, width=12cm,height=12cm}
\caption{The joint entropy of simple harmonic oscillator versus
          $\omega$ .}\label{eps3}
\end{figure}

\begin{figure}[htbp]
\centering \epsfig{file=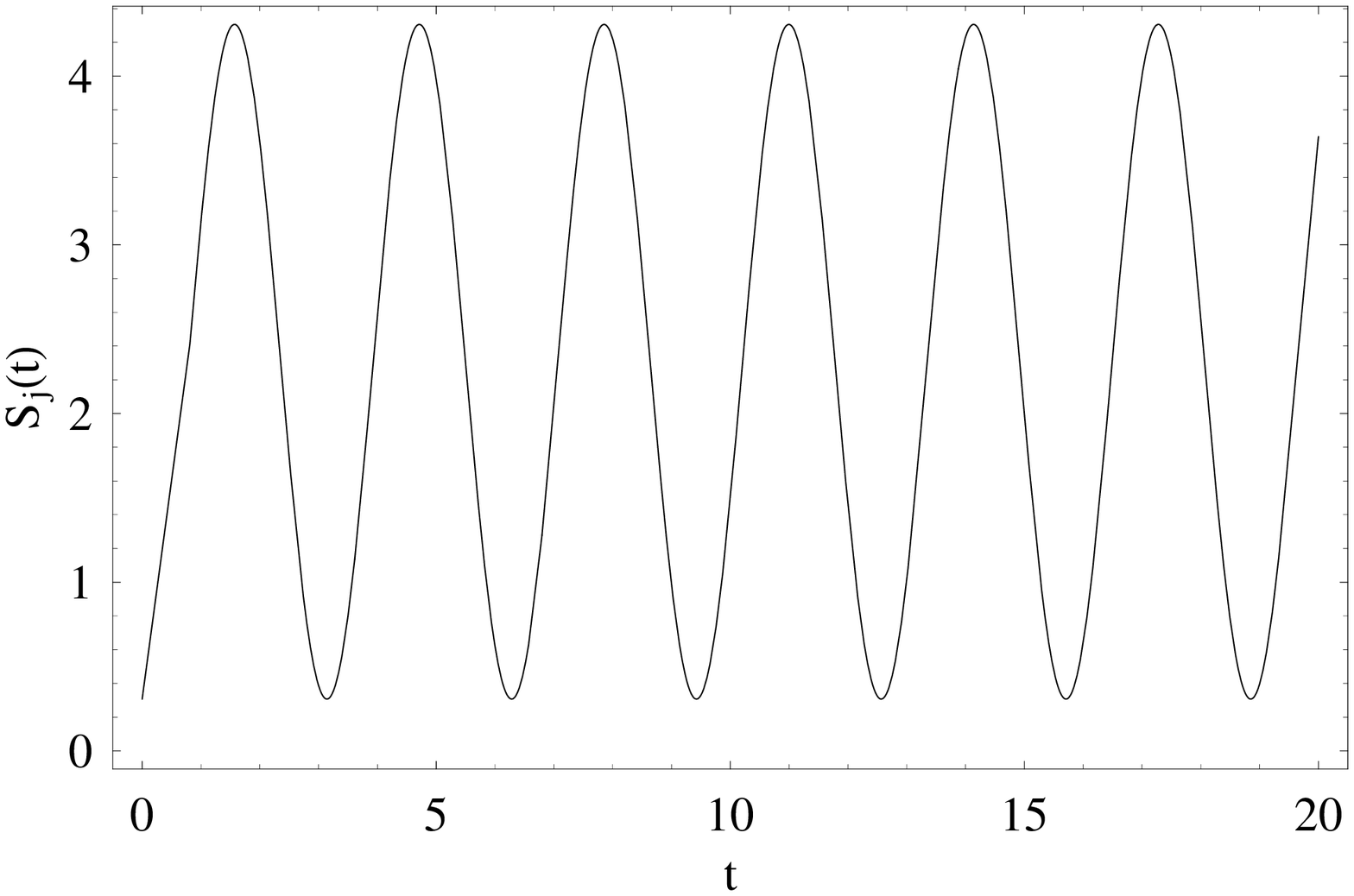, width=12cm,height=12cm}
\caption{The joint entropy of simple harmonic oscillator versus
          time.}\label{eps4}
\end{figure}

To get the path integral solution for the SHO, we must calculate its
action function. The Lagrangian of the system is given by
\begin{equation}
L(x,\dot{x},t)=\frac{m}{2}(\dot{x}^{2}-\frac{1}{2}\omega^{2}x^{2})
\end{equation}
Following a straightforward calculation, it is given by:
\begin{eqnarray}
    S(x_{cl}(t''),x_{cl}(t'))=\frac{m\omega}{2 \sin\omega t}[(x''^{2}_{cl}+x'^{2}_{cl})\cos\omega t-
    2 x'_{cl} x''_{cl}]
\end{eqnarray}
with $t=t''-t'$ and $x'_{cl}=x_0, x''_{cl}=x$. Substituting Eq. (9)
into Eq. (7), we obtain the Feynman kernel ~\cite{Feynman}:
\begin{eqnarray}
K(x,x_0;t)=(\frac{m\omega}{2\pi\hbar i\sin\omega
t})^\frac{1}{2}\exp\{-\frac{m\omega}{2i\hbar}[(x^{2}+x{_0}^{2})\cot\omega
t- \frac{2x_0 x}{\sin\omega t}]\}.
\end{eqnarray}
By the use of the Mehler-formula
\begin{eqnarray}
    e^{-(x^2+y^2)/2}\sum_{n=0}^{\infty}\frac{1}{n!}(\frac{z}{2})^2H_{n}(x)H_{n}(y)&=
    &\frac{1}{\sqrt{1-z^2}}\exp[\frac{4xyz-(x^2+y^2)(1+z^2)}{2(1-z^2)}]
\end{eqnarray}
where $H_{n}$ is Hermite polynomials, we can write the Feynman
kernel defining $x\equiv\sqrt{m\omega/\hbar}x_0$,
$y\equiv\sqrt{m\omega/\hbar}x$ and $z=e^{-i\omega T}$
\begin{equation}
K(x,x_0;t)=\sum_{n=0}^{\infty}e^{-itE_{n}/\hbar}\Psi^{*}(x_0)\Psi(x)
\end{equation}
with energy-spectrum and wave-functions:
\begin{equation}
    E_{n}=\hbar\omega(n+\frac{1}{2}),
\end{equation}
\begin{equation}
    \Psi_{n}(x)=(\frac{m\omega}{2^{2n}\pi\hbar
    n!^2})^\frac{1}{4}H_{n}(\sqrt{\frac{m\omega}{\hbar}}x)\exp(-\frac{m\omega}{2\hbar}x^2).
\end{equation}
Time dependent wave function of the SHO is defined as
\begin{equation}
    \Psi(x,t)=\int K(x,x_0;t)\Psi(x_0,0)dx_0.
\end{equation}
It can be written as
\begin{eqnarray}
    \Psi(x,t)=\Big(\frac{m\omega}{\pi
    \hbar}\Big)^{1/4}\exp\Big\{-\frac{\bar{\alpha}}{4}-\frac{\alpha^2}{2}-\frac{i\omega
    t}{2}\Big\}\exp\Big[-\frac{\bar{\alpha}^2}{4}e^{-2i\omega t}+
    \alpha\bar{\alpha}e^{-i\omega t}\Big]
\end{eqnarray}
where $\bar{x}$ or  $\bar{\alpha}$ is mean of the Gaussian curve.
The probability density has
\begin{equation}
|\Psi(x,t)|^{2}=\Big(\frac{m\omega}{\pi
    \hbar}\Big)^{1/2}\exp\Big[-(\alpha-\bar{\alpha}\cos\omega t)^{2}\Big]
\end{equation}
where $\alpha=\sqrt{\frac{m\omega}{\hbar}}x$. Thus it can be written
as
\begin{equation}
|\Psi(x,t)|^{2}=\Big(\frac{m\omega}{\pi
    \hbar}\Big)^{1/2}\exp\Big[-\frac{m\omega}{\hbar}(x-\bar{x}\cos\omega t)^{2}\Big]
\end{equation}
This has been shown in Fig.\ref{eps1}. In momentum space, the
probability density has the form
\begin{eqnarray}
|\tilde{\psi}(p,t)|^2=\Big(\frac{1}{m\omega\pi\hbar}\Big)^{1/2}\exp\Big[\frac{-p^2}{m\omega\hbar}+
\frac{m\omega\bar{x}^{2}}{2\hbar}\Big(\cos2\omega(t)-1\Big)-
\frac{2p\bar{x}}{\hbar}\sin\omega(t)\Big].
\end{eqnarray}
The joint entropy of harmonic oscillator becomes
\begin{equation}
    S_{j}(t)=\ln\frac{e}{2}+\frac{4m\omega}{\hbar}\bar{x}^{2}\sin^{2}\omega(t).
\end{equation}
In Fig.\ref{eps2}, the joint entropy of this system was plotted by
using Mathematica in three dimension. As known from fundamental
quantum mechanics and classical dynamics, displacement of simple
harmonic oscillator from equilibrium depends on harmonic functions
(e.g sine or cosine function). Therefore, other properties of the
SHO systems indicate the same harmonic behavior. If the frequency of
the SHO is sufficiently small, the system shows the same behavior as
the free particle\cite{Dunkel}. As seen from Fig.\ref{eps3} and
 Fig.\ref{eps4}, envelop of the sinusoidal curve is also
monotonically increase with omega and constant with time at constant
omega, respectively. When the frequency  increases, the joint
entropy of this system indicates a fluctuation with increasing
amplitude with time. If t goes to zero, it is important that Eq.(20)
is in agreement with following general inequality for the joint
entropy:
\begin{equation}
    S_{j}(t)\geq\ln(\frac{e}{2})
\end{equation}
originally derived by Leipnik for arbitrary one-dimensional
one-particle wave functions.
\subsection{Damped Harmonic Oscillator (DHO)}
The DHO is very important physical system in all physical systems
defining an interaction with its environment. The Lagrangian of the
DHO is given by
\begin{equation}
L(x,\dot{x},t)= e^{\gamma
t}\Big(\frac{m}{2}\dot{x}^{2}-\frac{m}{2}\omega^{2}x^{2}+j(t)x)\Big).
\end{equation}
Damped free particle kernel is
\begin{eqnarray}
K(x,t;x_0,0)=\Big(\frac{\gamma m e^{\gamma t/2}}{4\pi i \hbar \sinh
\frac{1}{2}\gamma t}\Big)^{2}\exp\Big(\frac{i\gamma m e^{\gamma
t/2}}{4\hbar \sinh \frac{1}{2}\gamma t}(x-x_0)^2\Big).
\end{eqnarray}
The DHO kernel has the form~\cite{um}
\begin{equation}
K(x,t;x_0,0)=\Big(\frac{m\omega e^{\gamma t/2}}{2\pi i \hbar \sinh
\omega t}\Big)^{1/2}\exp\Big(\frac{i}{\hbar}S_{cl}(x,x_0,t)\Big),
\end{equation}
or explicitly
\begin{eqnarray}
K(x,t;x_0,0)=\Big(\frac{m\omega e^{\gamma t/2}}{2\pi i \hbar \sin
\omega t}\Big)^{1/2}\exp\Big[\frac{im}{2\hbar}(ax^2+2bx_{0}^{2}+2x
x_{0}c+2xd+2x_{0}e-f)\Big].
\end{eqnarray}
Where the coefficients a, b, c, d, f are~\cite{um}
\begin{equation}
  a=(-\frac{\gamma}{2}+\omega\cot\omega t)e^{\gamma t},
\end{equation}
\begin{equation}
  b=(\frac{\gamma}{2}+\omega\cot\omega t),
\end{equation}
\begin{equation}
  c=(-\frac{\omega}{\sin\omega t}e^{\gamma t}),
\end{equation}
\begin{equation}
  d=\frac{e^{\gamma t}}{m\sin\omega t}\int_{0}^{t}j(t')e^{\gamma t'/2}\sin\omega t'dt',
\end{equation}
\begin{equation}
  e=\frac{1}{m\sin\omega t}\int_{0}^{t}j(t')e^{\gamma t'/2}\sin\omega (t-t')dt',
\end{equation}
\begin{eqnarray}
  f=\frac{1}{m^2\omega}\int_{0}^{t}\int_{0}^{t'}j(t')j(s)
   e^{\gamma (s+t'/2)}\sin\omega(t-t')\sin\omega s ds dt'.
\end{eqnarray}
The wave function $\psi_{n}(x,0)$ and energy eigenvalues become
\begin{eqnarray}
\psi_{n}(x,0)&=&N_{0}H_{n}(\alpha_{0}x)\exp\Big[-\frac{1}{2}\alpha_{0}x^2\Big]
\end{eqnarray}
and
\begin{equation}
    E_{n}=\Big(n+\frac{1}{2}\Big)\hbar\omega_{0}
\end{equation}
where $H_{n}(x)$ is the Hermite polynomial of order n and the
coefficients are
\begin{equation}
  \alpha_{0}=(\frac{m\omega}{\hbar})^{1/2},
  N_{0}=\frac{\alpha^{1/2}}{(2^nn!\sqrt{\pi})^{1/2}}.
\end{equation}
The time dependent wave function is obtained as~\cite{um}
\begin{eqnarray}
\psi_{n}(x,t)&=&\int_{-\infty}^{\infty}dx_{0}K(x,t;x_{0},0)\psi(x,0)\nonumber\\&=&N\frac{1}{(2^nn!)^{1/2}}
\exp\Big\{-i\Big[\Big(n+\frac{1}{2}\Big)\cot^{-1}\times\nonumber\\&\times&\Big(\frac{\gamma}{2\omega}+\cot\omega
t+f\Big)\Big]\Big\}\exp[-(Ax^2+\nonumber\\&+&2Bx)]H_{n}[D(x-E)].
\end{eqnarray}
To simplify the evaluation, we set $j(t)=0$. Such that kernel and
wave function of the DHO ~\cite{yeon} become
\begin{eqnarray}
K(x,t;x_0,0)&=&\Big(\frac{m\omega e^{\gamma t/2}}{2\pi i \hbar
\sin \omega
t}\Big)^{1/2}\exp\Big[\frac{im}{4\hbar}\Big(\gamma(x_{0}^{2}-e^{\gamma
t}x^2)+\frac{2\omega}{\sin\omega
t}\times\nonumber\\&\times&[(x_{0}^{2}+x^2e^{\gamma t})\cos\omega
t-2e^{\gamma t/2}xx_0]\Big)\Big]
\end{eqnarray}
where $\omega=(\omega_{0}^2-\gamma^2/4)^{1/2}$ and
\begin{eqnarray}
\psi_{n}(x,t)=\frac{N}{(2^nn!)^{1/2}}
\exp\Big\{-i\Big[\Big(n+\frac{1}{2}\Big)\cot^{-1}\Big(\frac{\gamma}{2\omega}+\cot\omega
t\Big)\Big]\Big\}H_{n}[Dx]\exp[-Ax^2].
\end{eqnarray}
Where D, A and N are
\begin{equation}
    D(t)=\frac{\alpha e^{\gamma t/2}}{\eta(t)\sin\omega t},
\end{equation}
\begin{equation}
    \eta^2(t)=\frac{\gamma^2}{4\omega^2}+\frac{\gamma}{\omega}\cos\omega
    t+\csc^2\omega t,
\end{equation}
\begin{eqnarray}
A(t)=\frac{m\omega}{2\hbar}e^{\gamma
t}\Big[\frac{1}{\eta^2(t)\sin^2\omega
t}+i\Big(\frac{\gamma}{2\omega}-\cot\omega
t+\frac{\gamma/2\omega+\cot\omega t}{\eta^2\sin^2\omega
t}\Big)\Big],
\end{eqnarray}
and
\begin{equation}
N(t)=\Big(\frac{m\omega}{\pi\hbar}\Big)^{1/4}\frac{\exp(\frac{\gamma
t}{4})}{\eta(t)(\sin\omega t)^{1/2}}.
\end{equation}

\begin{figure}[htbp]
\centering \epsfig{file=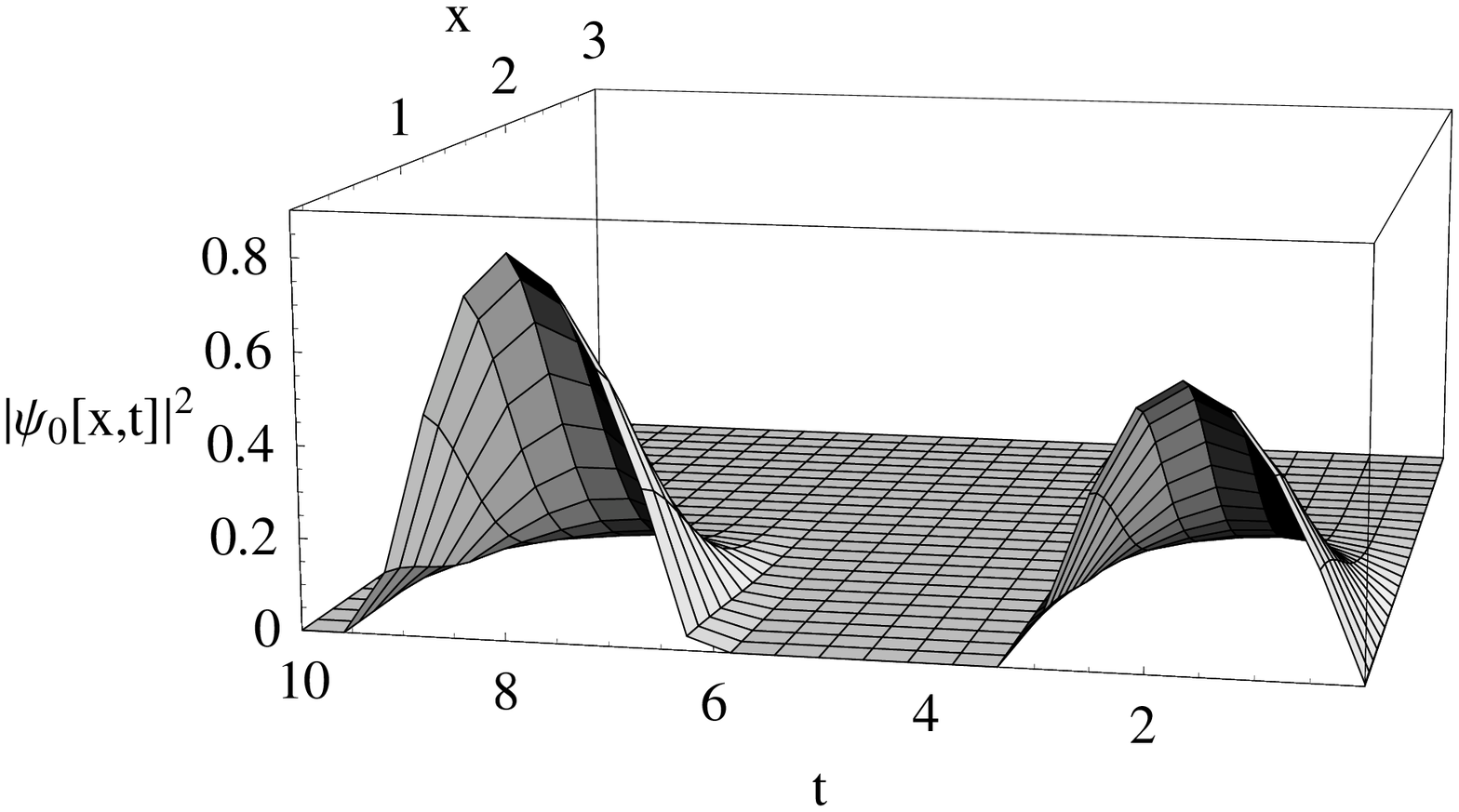, width=12cm,height=12cm}
\caption{The probability function as a function of time and
          coordinate at $\gamma=0.1$.}\label{eps5}
\end{figure}

\begin{figure}[htbp]
\centering \epsfig{file=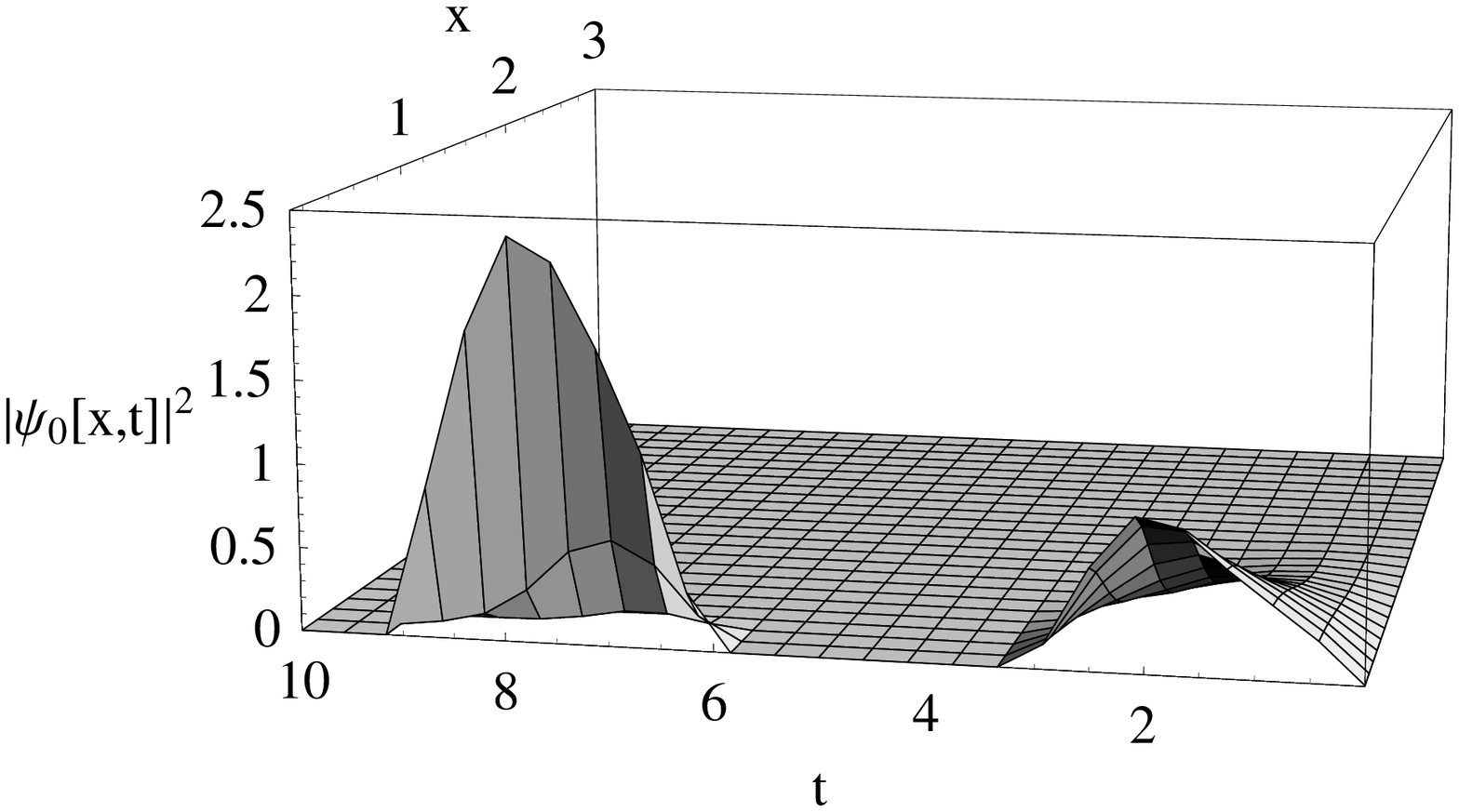, width=12cm,height=12cm}
\caption{The probability function as a function of time and
          coordinate at $\gamma =0.5$. }\label{eps6}
\end{figure}

The ground state wave function is given by
\begin{eqnarray}
\psi_{0}(x,t)=N(t)
\exp\Big\{-i\Big[\Big(\frac{1}{2}\Big)\cot^{-1}\Big(\frac{\gamma}{2\omega}+\cot\omega
t\Big)\Big]\Big\}\exp[-A(t)x^2].
\end{eqnarray}
So the probability distribution in coordinate space becomes
\begin{eqnarray}
|\psi_{0}(x,t)|^2&=&N(t)^{2}\exp[-2A'(t)x^2]
\end{eqnarray}
where $A'$ is defined by
\begin{eqnarray}
A'(t)&=&\frac{m\omega}{2\hbar}e^{\gamma
t}\Big[\frac{1}{\eta^2(t)\sin^2\omega t}\Big].
\end{eqnarray}
The probability density in coordinate space is shown in
Fig.\ref{eps5} and Fig.\ref{eps6} for the different values of
$\gamma$. The probability density in momentum space can be written
easily
\begin{eqnarray}
|\psi_{0}(p,t)|^2&=&\frac{N(t)^{2}}{\sqrt{2A(t)A(t)^\dag\hbar}}\exp\Big[-\frac{p^2}{2\hbar^2
}\frac{A'(t)}{A(t)A(t)^\dag}\Big].
\end{eqnarray}
The time dependent joint entropy can be obtained from Eq. (2) as
\begin{eqnarray}
S_{j} (t)=N(t)^2\sqrt{\frac{\pi}{2A'(t)}}\Big[(\ln
N(t)^2-\frac{1}{2})-\frac{1}{2}\sqrt{\frac{1}{2A(t)
A(t)^\dag}}\Big(\ln\frac{ N(t)^2}{2A(t)
A(t)^\dag}-\frac{1}{2}\Big)\Big]-\ln2\pi.
\end{eqnarray}

\begin{figure}[htbp]
\centering \epsfig{file=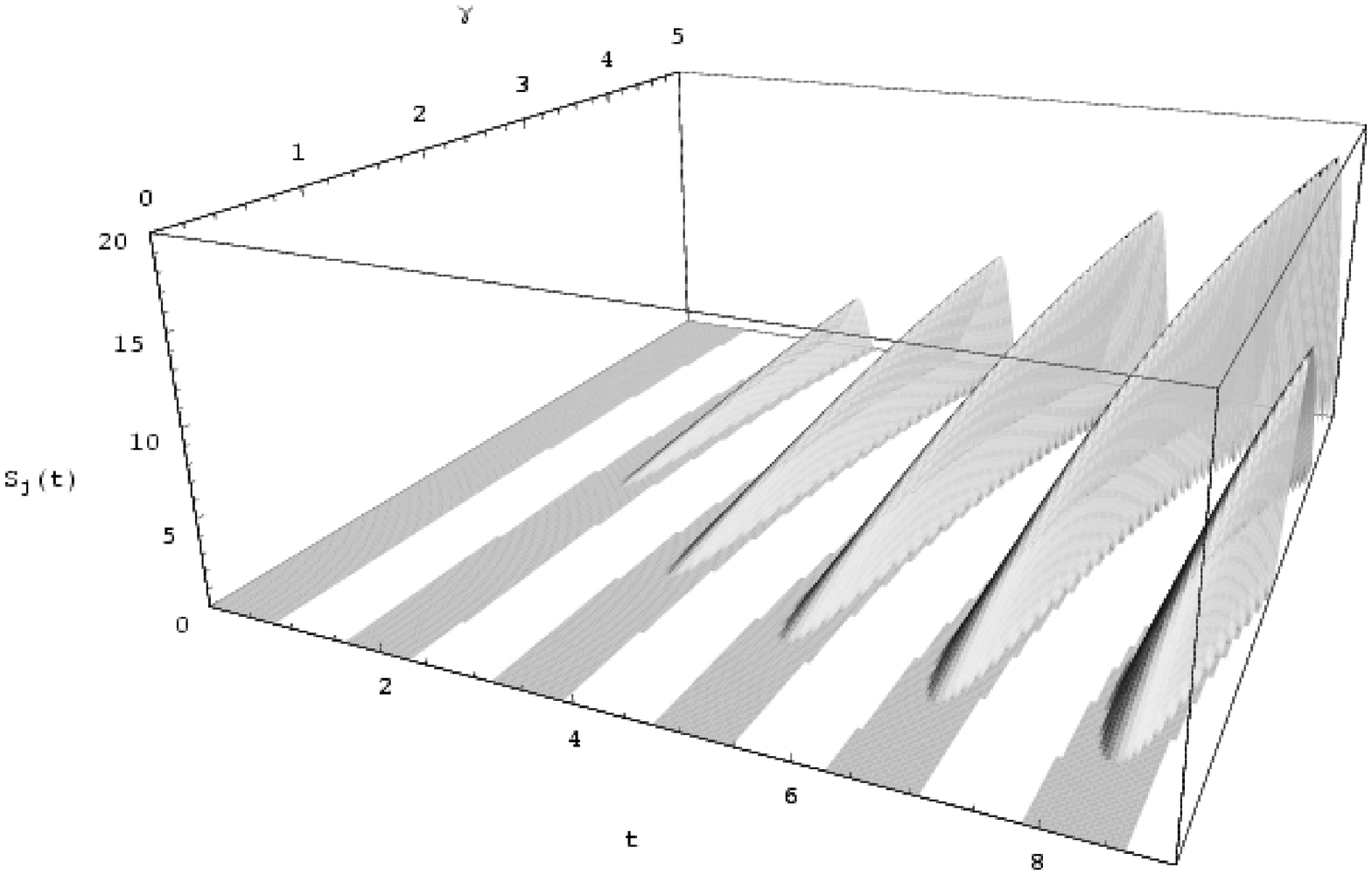, width=12cm,height=12cm}
\caption{The 3D graph of the joint entropy of damped harmonic
          oscillator for damping factor$(\gamma)$ at $\omega_0=2$. }\label{eps7}
\end{figure}

\begin{figure}[htbp]
\centering \epsfig{file=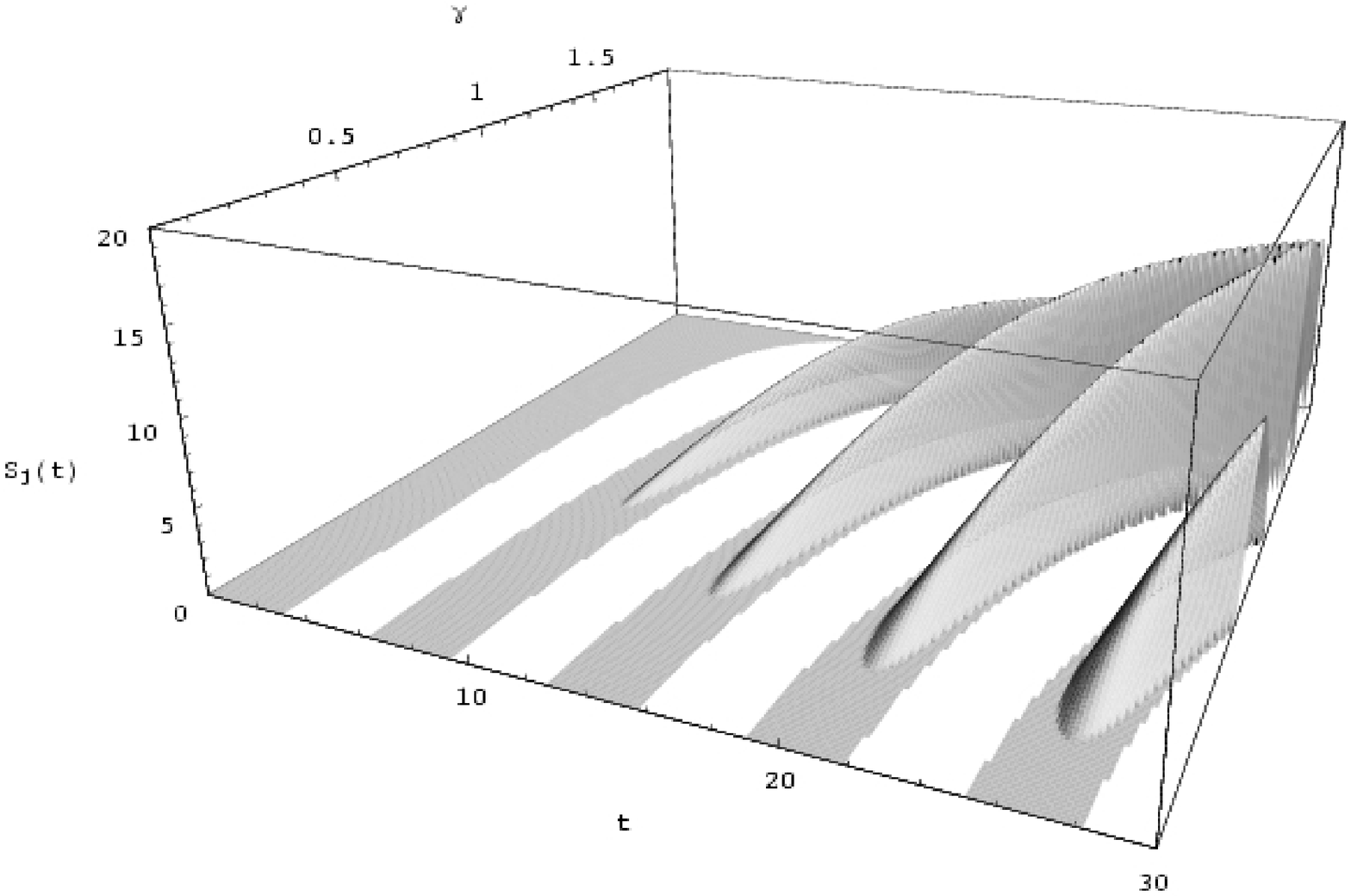, width=12cm,height=12cm}
\caption{The 3D graph of the joint entropy of damped harmonic
          oscillator for damping factor$(\gamma)$ at $\omega_0=1$.}\label{eps8}
\end{figure}

The joint entropy depends on damping factor $\gamma$. When
$\gamma\rightarrow 0$, all the above results are converged to
simple harmonic oscillator. However, when the $\gamma\neq 0$, the
joint entropy has remarkably different features of the SHO. As can
be seen in Fig.\ref{eps7} and Fig.\ref{eps8}, the joint entropy of
the DHO has very interesting properties. One of the most important
properties of the joint entropy is the probability  of taking
 values for small $\gamma$ values. As we know from
literature the joint entropy must be positive and monotonically
increase.However, this system has different properties from
literature because of periodically discontinuity of the joint
entropy. On the other hand, envelop of this curve is also
monotonically increase with time for large $\gamma$. As can be
shown these results, the envelop of the joint entropy curves has
general properties as monotonically increase for quantum systems.
Thus, we have found that the joint entropy is depend on properties
of investigated system.

\section{Conclusion}
We have investigated the joint entropy for explicit time dependent
solution of one-dimensional harmonic oscillators. We have obtained
the time dependent wave function by means of Feynmann Path integral
technique. Our results show that in the simple harmonic oscillator
case, the joint entropy fluctuated with time and frequency. This
result indicates that the information periodically transfer between
harmonic oscillators.

On the other hand, in the DHO case, the joint entropy shows a
remarkable smooth discontinuities with time. It also depends on
choice of initial values of parameter i.e. $\omega$. These results
can be explained as the information exchange between harmonic
oscillator and system which is supplied damping. But the
information exchange appears in certain values of time for
damping. If the damping factor increases, the information entropy
has not periodicity anymore. Moreover, for certain values of the
damping factor, the transfer of information between systems is
exhausted.

\section{Acknowledgements}

This research was partially supported by the Scientific and
Technological Research Council of Turkey.

\newpage
\newpage
\newpage
\newpage
\newpage
\newpage
\newpage
\newpage
\end{document}